\documentclass[proof]{WileyASNA-v1}

\articletype{Article Type}%

\received{31 August 2018}
\accepted{5 October 2018}

\begin{document}

\title{High Energy Emission and Its Variability in Young Stellar Objects}

\author[1,2]{Costanza Argiroffi}

\authormark{C. Argiroffi}

\address[1]{\orgname{University of Palermo}, \orgdiv{Department of Physics and Chemistry}, \orgaddress{\state{Piazza del Parlamento 1, 90134, Palermo}, \country{Italy}}}

\address[2]{\orgdiv{INAF - Osservatorio Astronomico di Palermo}, \orgname{ }, \orgaddress{\state{Piazza del Parlamento 1, 90134, Palermo}, \country{Italy}}}

\corres{*Costanza Argiroffi, \email{costanza.argiroffi@unipa.it}}

\abstract{Young stars show a variety of highly energetic phenomena, from accretion and outflow processes to hot coronal plasmas confined in their outer atmosphere, all regulated by the intense stellar magnetic fields. Many aspects on each of these phenomena are debated, but, most notably, their complex mutual interaction remains obscure.
In this work I report how these phenomena are simultaneously responsible for the high-energy emission from young stars, with a special focus on the expected and observed variability in the X-ray band.
Investigating variations in the X-ray emission from young stars allows us to pose constraints on flare and coronal plasma properties, coronal heating, accretion stream properties, and accretion geometries. All these results are important building blocks for constructing a comprehensive picture of the complex magnetosphere of young stars.}


\maketitle

\section{Introduction}
\label{intro}


Stars form in dense embedded prestellar cores, results of the gravitational collapse of interstellar clouds. In its pre-main sequence life, a newly born star is still contracting its volume, and powers itself with the gravitational energy released \citep[e.g.][]{Hartmann2002,Larson2003}. During this phase the photospheric temperature remains approximately constant, while the stellar luminosity decreases. When the internal temperature is high enough to make the hydrogen burning in the stellar nucleus effective, the young star ends its contraction and enters in the main sequence (MS) phase \citep{SiessDufour2000,BaraffeHomeier2015}. Simultaneously to the evolution of the central star, the circumstellar material, remnant of the prestellar core, evolves as well. The young star is first surrounded by a thick envelope, that then settles into a circumstellar disk, from which the star continues to accrete material. On a time scales of a few Myr the accretion phase ends, and the remnant of the disk is eventually dispersed and/or translated into a planetary system \citep{Armitage2011,WilliamsCieza2011}.


Even if star formation occurs in cold environments, young stars are source of hard emission, because of highly energetic processes occurring in their atmospheres and circumstellar environments \citep{FeigelsonMontmerle1999}. All these highly energetic phenomena are generated and regulated by the intense stellar magnetic fields. Young low-mass stars ($M\lesssim2M_{\odot}$) are in fact rapid rotators \citep[$P_{\rm rot}\sim1-10$\,d, e.g.][]{StassunMathieu1999,AfferMicela2013}, and have convective envelopes under their photosphere. Dynamo processes effectively operate, generating magnetic fields of a few kG \citep{JohnsKrull2007} since stellar birth. These strong magnetic fields regulate accretion and outflows processes, allow the outward transport of angular momentum, and are responsible for heating and confinement of hot coronal plasmas located in the outer stellar atmosphere. All these phenomena are source of hard and variable emission in young stars. Most notably, all these phenomena interact among themselves. How this interplay occurs is unclear, and different interaction mechanisms have been proposed.


In this review I focus on the processes responsible for the X-ray emission in young low-mass stars. In particular, I show that investigating X-ray variability is a powerful means to understand properties and geometries of young stellar magnetospheres. I consider in this work only two X-ray emission mechanisms: coronal plasma and plasma heated in the mass accretion process. Note however that X-rays can also originate from other processes. X-rays can be emitted by the circumstellar disk, where  cold Fe atoms, exposed to the stellar X-ray emission, are responsible for the fluorescent K$\alpha$ emission line at 6.4\,keV \citep{ImanishiKoyama2001,TsujimotoFeigelson2005,GiardinoFavata2007}. The properties and diagnostics provided by this emission line are discussed in the S.~Sciortino contribution in this volume. Finally, soft X-rays are sometimes observed also from outflows and collimated jets  of very young accreting stars \citep{PravdoFeigelson2001,FavataFridlund2002}. This emission originates in shocks occurring between the ejected blobs and between the blobs and the ambient medium \citep{BonitoOrlando2010}. This emission, because of its extremely low signal-to-noise ratio, appears stable within individual observations, and between observations, apart for small proper motion effects \citep{BonitoOrlando2011}.


Different nomenclatures are used in referring to young stars. In general, irrespective of their mass and/or evolutionary phase, young stars are named as pre-main sequence (PMS) stars or young stellar objects. Considering the evolution of the circumstellar material, young stars are classified as: Class~0, when the source is still its protostellar phase; Class~I, when the star is formed but it is still surrounded by a dense envelope, Class~II when the circumstellar envelope is evolved into a disk; Class~III, when the circumstellar disk is finally dispersed. Finally, young low-mass ($M\lesssim2\,M_{\odot}$) stars are usually called T~Tauri stars, distinguishing between classical T~Tauri stars, when the accretion process is still at work, and weak-line T~Tauri stars, when the accretion process has already ended. 

In Section~\ref{corona} I report the main properties of coronal emission from young low-mass stars, presenting its typical variability, and the insights it provides. Section~\ref{accretion} is dedicated to the accretion-driven X-ray emission, discussing the predicted and observed variability on different time scales. Section~\ref{rotmod} is dedicated to the results obtained from the observed rotational modulation effects.  In Section~\ref{cor_accr_interplay} I briefly report on the evidence obtained studying X-ray variability on the possible interplay between the corona and the accretion process. 

\vspace{-2mm}
\section{Coronal emission}
\label{corona}


T~Tauri stars possess hot coronal plasmas in their outer atmosphere. The physics of these coronal plasmas is thought to be analogous to that of late-type MS stars: stellar internal motions due to rotation and convection produce strong magnetic fields, via dynamo mechanisms, that are the responsible for the heating and confinement of these coronal plasmas \citep{NoyesHartmann1984}. Coronal plasmas of T~Tauri stars have temperatures of $\sim10-100$\,MK \citep[e.g.][]{ArgiroffiFavata2006,FranciosiniPillitteri2007} being therefore bright in the X-ray band, with $L_{\rm X}/L_{\rm bol}$ up to $10^{-3}$ \citep{PreibischKim2005,WolkHarnden2005,BriggsGuedel2007}.


Magnetic activity of MS stars is well described by the Rossby number $R_o$, the ratio between the rotational period and the convective turnover time. Depending on $R_o$, MS stars display different activity regimes: the non-saturated regime for $R_o\gtrsim0.13$, with $L_{\rm X}/L_{\rm bol}$ decreasing for increasing $R_o$; the saturated regime for $R_o\lesssim0.13$, with $L_{\rm X}/L_{\rm bol}\approx10^{-3}$ \citep{DobsonRadick1989,PizzolatoMaggio2003,WrightDrake2011}. Evidence of a third regime, the supersaturation, exists: extremely rapid rotators have $L_{\rm X}/L_{\rm bol}$ reduced with respect to saturation values \citep{ArgiroffiCaramazza2016}.


Even if the coronae of T~Tauri stars appear similar to that of MS stars (in terms of temperatures and $L_{\rm X}/L_{\rm bol}$ values), their activity levels do not seem to depend on $R_o$ \citep{PreibischKim2005,WolkHarnden2005,BriggsGuedel2007}. Evidence of $L_{\rm X}/L_{\rm bol}$ vs $R_o$ patterns seem to start at ages of $\sim13$\,Myr \citep{ArgiroffiCaramazza2016}. This difference between PMS and MS stars could be due to different magnetic properties because of the different internal structures and/or different velocity fields in stellar interior. In addition, the activity level of T~Tauri stars was found to depend on the accretion status, with accreting stars displaying on average activity levels lower than that of non-accreting stars \citep{FlaccomioMicela2003,PreibischKim2005}, suggesting that accretion affects coronal activity.

\begin{figure}
\centering
\includegraphics[width=0.45\textwidth]{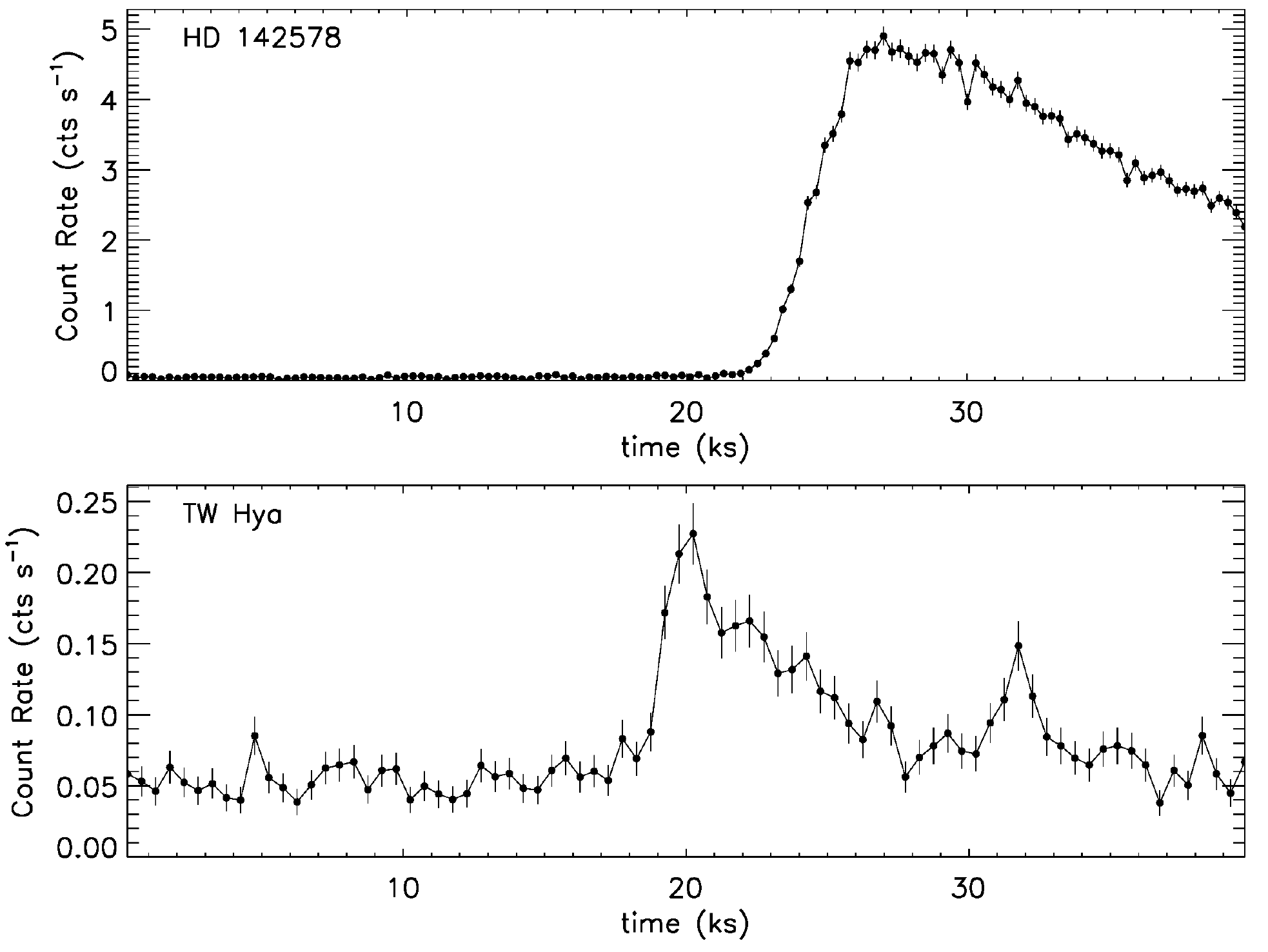}
\caption{X-ray light curves of two young stars showing strong flares. HD 142578 is a 5\,Myr member of the Upper Scorpius association, and its flare was registered with XMM/PN \citep[ObsID~0112380101,][]{ArgiroffiFavata2006}. TW~Hya is an 8\,Myr old accreting star, and this flare was registered with Chandra/HETGS \citep[ObsID~7437,][]{BrickhouseCranmer2010} \label{fig1}}
\end{figure}


Coronal emission from T~Tauri star is variable at almost all the explored time scales (from minutes to years). Short time-scale ($\sim1-100$\,ks) variability is mainly due to flares \citep{MontmerleKochMiramond1983,FeigelsonBroos2002,PreibischZinnecker2002}, rapid releases of magnetic energy observed as X-ray emission bursts (Fig.~\ref{fig1}). Variability is observed also on time scales of days, because of stellar rotation (see Section~\ref{cor_accr_interplay}). Finally young stars show $L_{\rm X}$ variable also on time scales of years \citep{ArgiroffiFavata2006,CaballeroAlbaceteColombo2010}, possibly due to coronal X-ray activity cycle.


By investigating X-ray emission of strong and isolated flares it is possible to infer important properties of the involved coronal structures. Flares originate from rapid releases of magnetic energy occurring in the outer stellar atmospheres. The released energy heats chromospheric material, that first expands upward filling the above coronal structures on time scales of a few ks, and then radiatively cools down on time scales of $10-100$\,ks. The flare emission in the X-ray band comes from the coronal portion of the flaring loop, where the hottest plasma, with temperatures up to $\sim100$\,MK, is located.


Different methods, based on inspecting the flare rising phase, the flare decay phase, or the flare oscillations, allow us to infer the flaring loop length \citep{RealeBetta1997,Reale2007,LopezSantiagoCrespoChacon2016}. T~Tauri stars usually show compact flaring structures, with loop semi-length $L\lesssim R_{\star}$ \citep{PillitteriMicela2005,ArgiroffiFavata2006,FranciosiniPillitteri2007}, analogous to that observed in the solar corona. In addition to these normal loops, young stars sometimes host extremely long flaring loops, with $L\gtrsim5-10\,R_{\star}$ \citep{FavataFlaccomio2005,GiardinoFavata2007,GetmanFeigelson2008a,McClearyWolk2011,ArgiroffiFlaccomio2011,LopezSantiagoCrespoChacon2016,RealeLopezSantiago2018}. Such very long loops have been found only in young stars, indicating that extended coronal structures are typical in this early evolutionary phase. It has been argued \citep[e.g.][]{FavataFlaccomio2005,GetmanFeigelson2008b} that such long corona loops, in case of accreting stars, could be loops connecting the stellar photosphere to the accretion disk, allowing a direct interplay between coronal activity and accretion process (see Section~\ref{cor_accr_interplay}).


The presence of extended coronal structures in young stars fits with the observed supersaturation regime of stellar activity. Supersatuation can be explained assuming that centrifugal forces disrupt the most extended coronal structures, limiting the amount of coronal plasma \citep{JardineUnruh1999}. The observed pattern of supersaturation indicates that young stellar coronae  have loops with $L$ up to $\sim3R_{\star}$ \citep{ArgiroffiCaramazza2016}.


Time-resolved X-ray spectroscopy of individual flares provides constraints also on the so-called First Ionization Potential (FIP) effect in young stars. Coronal and photospheric metallicities of active stars are found to be different. In particular, coronal plasma of active stars is usually depleted of metals, especially that with low FIP. In flaring loops of young stars abundances sometimes are significantly enhanced with respect to that of the quiescent corona \citep[e.g.][]{ImanishiTsujimoto2002,ArgiroffiFavata2006}.
Assuming that the mechanism responsible for the FIP effect operates on time scales longer than flare duration, then the quiescent corona is expected to be depleted of metals, while a flaring loop, filled of material just evaporated from the underlying photosphere, should reveal a different metallicity, as indeed observed.


Strong flare emission from young stars is relevant also for source detection. The quiescent X-ray emission of young stars (maybe because still embedded in dense envelopes heavily absorbing their radiation) could be too weak to be detected. The transient emission can instead provide a signal-to-noise ratio high enough to allow the source detection in short time intervals \citep{PizzocaroStelzer2016}.


In addition to the study of individual flares, constraining the flare energy distribution allows us to test the heating mechanisms of stellar coronae, and in particular the hypothesis that coronal plasma  (in T~Tauri stars as well as in all late-type active stars) is entirely heated by flares, from nano-flare (radiated X-ray energy $E_{\rm X}\sim10^{23}-10^{25}$\,erg) to major flares (with $E_{\rm X}$ up to $\sim10^{37}$\, erg), by analogy with the scenario proposed studying the solar corona \citep{Hudson1991}. The number of flares observed at different energy is well described by the power law ${\rm d} N/{\rm d} E \propto E^{-\alpha}$, similar to that observed for the Sun and for other active stars. In young stars, the inferred $\alpha$ values range from 1.5 to 2.5 \citep{WolkHarnden2005,StelzerFlaccomio2007,CaramazzaFlaccomio2007,AlbaceteColomboCaramazza2007,CaramazzaMicela2012}, supporting indeed the scenario of coronal plasmas entirely heated by flares.


Constraining the flare energy distribution is essential also because flares are often used as proxies of associated phenomena, non directly observable in stars. This is the case of stellar coronal mass ejections (CME). In the case of young stars inferring the amount of energy and mass extracted by CME is important to understand young-star evolution, the evolution of stellar rotation, and the evolution of circumstellar disks and newly formed planetary systems \citep{OstenWolk2015}.

\vspace{-2mm}
\section{Accretion process}
\label{accretion}

During the first Myr of their life, low-mass stars accrete material from their circumstellar disks. The accretion process is regulated by the stellar magnetic field. The magnetospheric accretion scenario predicts that the stellar magnetic field is strong enough to disrupt the inner disk at a few stellar radii. The inner disk material is loaded into the magnetic flux tubes, and moves along the field line toward the star, forming an accretion stream \citep[e.g.][]{Koenigl1991,RomanovaOwocki2015,HartmannHerczeg2016}.

\begin{figure}
\centering
\includegraphics[width=0.45\textwidth]{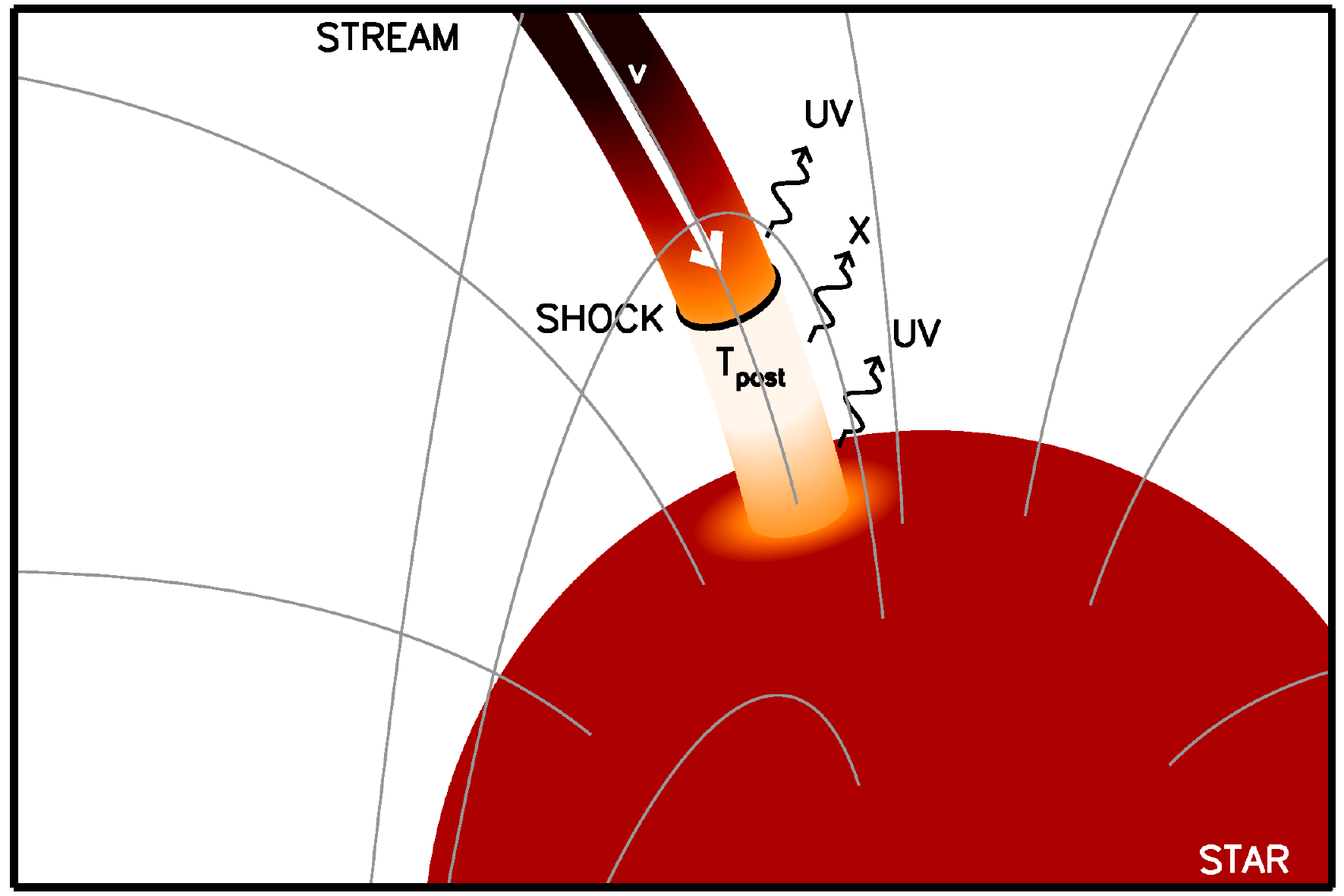}
\caption{Cartoon of the accretion-shock region in a young accreting star. \label{fig2}}
\end{figure}

The accreting material is expected to free fall toward the star, arriving at the base of the accretion stream with velocities of $\sim300-500\,{\rm km\,s^{-1}}$. There, because of the impact with the denser stellar atmosphere, accreting material forms a strong shock (Fig.~\ref{fig2}), that heats the infalling material up to temperatures are of a few MK, and makes the shock region a source of soft X-rays \citep{Ulrich1976,Gullbring1994}.

The plasma component heated in the accretion shock has been detected and identified thanks to the measure of its density \citep[$n_{\rm e}\sim10^{11}-10^{13}\,{\rm cm^{-3}}$,][]{KastnerHuenemoerder2002,SchmittFavata2005,GuntherLiefke2006,ArgiroffiMaggio2007}, significantly higher than that of coronal plasma. While density, temperature, and velocity of the post-shock material  \citep{ArgiroffiDrake2017} agree with model predictions \citep{SaccoArgiroffi2008}, the amount of soft X-rays emerging from the post-shock is significantly lower than that expected \citep{CurranArgiroffi2011}. Possibly, the major part of X-ray emission is locally absorbed, because of the surrounding stellar atmosphere and pre-shock material \citep{Drake2005,SaccoOrlando2010}. This local absorption causes the heating of these nearby regions, hence contributing in forming a hot spot in the stellar atmosphere, with  $T\sim10^{4}$\,K, and evidenced by excess emission in the optical and UV bands.

Focusing on the high energy emission emerging from the post-shock, variability is expected on different time scales, and for different phenomena.

HD and MHD models predict that the post-shock region is not stable, but it should oscillate quasi periodically because of thermal instabilities \citep{SaccoArgiroffi2008,KoldobaUstyugova2008,SaccoOrlando2010}. The period of these oscillations depends on the accretion stream density, and ranges from $10^{-2}$ to $ 10^{4}$\,s . Observed emission in soft X-rays and far UV from young accreting stars does not show any periodicity \citep{DrakeRatzlaff2009,GuentherLewandowska2010}. Note however that the accretion stream is likely composed of different fibrils with different densities, that remain separated because of the strong magnetic field. Each fibrils is then expected to oscillate with its own period and phase, preventing the presence of any global oscillation in the X-rays emerging from the post-shock.

Irrespective of oscillations and local absorption effects, the average post-shock X-ray luminosity $L_{\rm X\,accr}$ is expected to correlate with the mass accretion rate $\dot{M}$ \citep{SaccoOrlando2010}. $\dot{M}$ is observed to vary from time scales of hours to years. The possibility to detect correlated variations in $\dot{M}$ and $L_{\rm X\,accr}$ is complicated because variations in $\dot{M}$ means changes in density and/or cross section and/or numbers of the accretion streams. As a consequence also geometry and location of the post-shock region, and hence of local absorption, are expected to change as well. There have been a few attempts to investigate this issue. During a long X-ray monitoring of the young accreting star TW~Hya a significant decrease in the local absorption (by a factor of 3 on time scales of $\sim3$ \,d) was interpreted it in terms of changes in streams geometry and properties, corresponding to a decrease in $\dot{M}$ by a factor of $\sim5$ \citep{BrickhouseCranmer2012}. Evidence of variations in the post-shock plasma properties, due to intrinsic variations in the accretion stream, has been obtained for the young accreting star V4046~Sgr, where the post-shock density decreased by a factor of 10 on a time scale of 1\,d \citep{ArgiroffiMaggio2014}. Finally, simultaneous increases in soft X-rays (probing the post-shock plasma) and optical (probing the hot accretion spot) have been observed in young accreting stars of NGC~2264, and interpreted in terms of accretion bursts \citep{GuarcelloFlaccomio2017}.

\vspace{-2mm}
\section{Rotational modulation}
\label{rotmod}

X-ray emission from young stars is variable also because of stellar rotation. Rotational modulation is expected for both coronal and shock-heated plasma. Both these components, being linked to the magnetic field, have non-homogeneous distributions on the stellar surface. In stars with accretion, in addition to modulation due to stellar occultation, eclipses of X-rays can also be due to circumstellar structures, like accretion streams and inner disk warps. Especially in these cases, detection of rotational modulation is a powerful means to constrain the intricate accretion geometry.

During the long X-ray monitoring of the Orion Nebula Cluster, rotational modulation of X-ray coronal emission has been observed in some stars. In these cases X-rays were observed to vary with a period $P_{\rm X}$ that in some cases was $\sim P_{\rm rot}$, as expected, and in other cases was $\sim 0.5P_{\rm rot}$ \citep{FlaccomioMicela2005}. Models of spatial distribution of coronal plasmas in young stars, obtained starting from stellar surface magnetograms, confirm that X-ray rotational modulation is expected, and suggest that the two different periods observed can be explained assuming different stellar inclinations with respect to us, and in particular stars with $i\sim90^{\circ}$ would favor $P_{\rm X}\sim 0.5 P_{\rm rot}$, while stars with $i\sim 30^{\circ}-60^{\circ}$ would favor $P_{\rm X}\sim P_{\rm rot}$ \citep{GregoryJardine2006}.

In some young accreting stars, the observed soft X-rays showed variations compatible with transient increases in the absorption, likely due to accretion streams and/or disk warps periodically passing along the line of sight \citep{SchmittRobrade2007,FlaccomioMicela2010}.  

Similarly, variations due to rotational modulation have been observed also for the accretion-driven X-rays. In the young accreting star V2129~Oph, the plasma heated in the accretion shock disappears during stellar rotation likely because of the passage of the accretion stream across the line of sight to the shock region \citep{ArgiroffiFlaccomio2011}. The young accreting star V4046~Sgr, monitored in X-rays for more than two complete rotations, showed that the soft X-rays emerging form the accretion shock region display periodic variations, with an X-ray period again corresponding to $0.5\,P_{\rm rot}$ \citep{ArgiroffiMaggio2012}. This result, previously obtained for coronal plasma, indicates that also material heated in the accretion shock possibly follows an analogous spatial distribution on the stellar surface.

\vspace{-2mm}
\section{Interactions between accretion and corona}
\label{cor_accr_interplay}

All the high energy processes at work in young accreting stars are thought to interact among themselves. Different possible interaction mechanisms have been proposed. Firstly all these high-energy processes are related to the magnetic fields, hence the magnetosphere properties and variations should simultaneously affect all of them \citep[e.g.][]{JohnstoneJardine2014}. Moreover, all these high-energy processes produce ionizing radiation, affecting the coupling between cold matters and magnetic fields. Therefore, by this mechanism, they could induce feedback effects on their efficiency \citep[e.g.][]{DrakeErcolano2009}. Finally, all these processes cause angular momentum loss, hence influence stellar rotation, and eventually the process responsible for the magnetic field generation \citep[e.g.][]{FlaccomioDamiani2003}.

The existence of long coronal loops, as inferred from flare analysis, suggests that the presence of the disk favors the formation of long coronal loops extending from the inner disk to the stellar photosphere \citep{FavataFlaccomio2005}. This scenario, if confirmed, would indicate that the accretion process does directly affect coronal plasmas, and that the different $L_{\rm X}/L_{\rm bol}$ values observed for accreting and non-accreting stars could be explained by this kind of interaction mechamisms.


The existence of star-disk coronal loops could generate a new class of flares. In fact, magnetic reconnections could occur also near the disk footpoint of the loop. That would allow the release of magnetic energy and the trigger of flares near the disk. MHD simulations of this scenario proved that such flares would strongly perturb the local portions of the disk, and could eventually trigger accretion bursts onto the star \citep{OrlandoReale2011}.

\vspace{-2mm}
\section{Conclusions}

Young stars are source of high energy emission, mainly from coronal plasma and accretion process. Inspecting and understanding this X-ray emission is essential to understand the physics of young stars.

\bibliographystyle{Wiley-ASNA}
\bibliography{argiroffi}

\end{document}